\documentclass[prl,twocolumn]{revtex4-1}

\usepackage[english]{babel}
\usepackage{graphicx}
\usepackage{amssymb}
\usepackage[T1]{fontenc}
\usepackage[applemac]{inputenc}
\usepackage{textcomp}
\usepackage{epsfig}
\usepackage{mathrsfs}
\usepackage{float}
\usepackage{tikz}

\def\comment#1{}
\newcommand{\Tr}{\mbox{Tr\,}}

\newcommand{\beg}{\begin{eqnarray}}
\newcommand{\eee}{\end{eqnarray}}

\def\cm#1{}

\newcommand{\be}{\begin{equation}}
\newcommand{\ee}{\end{equation}}
\newcommand{\ba}{\begin{eqnarray}}
\newcommand{\ea}{\end{eqnarray}}
\newcommand{\beq}{\begin{equation}}
\newcommand{\eeq}{\end{equation}}
\newcommand{\bea}{\begin{eqnarray}}
\newcommand{\eea}{\end{eqnarray}}
\newcommand{\bastar}{\begin{eqnarray*}}
\newcommand{\eastar}{\end{eqnarray*}}

\newcommand{\ignore}[1]{}

\begin{document}

\title{Diagrammatic Monte Carlo procedure for the spin-charge transformed Hubbard model}
\author{Johan~Carlstr\"om}
\affiliation{
Department of Physics, Stockholm University, 106 91 Stockholm, Sweden
}
\date{\today}
\begin{abstract}
Using a dual representation of lattice fermion models that is based on spin-charge transformation and fermionisation of the original description, I derive an algorithm for diagrammatic Monte Carlo simulation of strongly correlated systems. 
This scheme allows eliminating large expansion parameters, as well as large corrections to the density matrix that generally prevent diagrammatic methods from being efficient in this regime. 
As an example, I compute the filling factor for the Hubbard model at infinite onsite repulsion and compare the results to controllable data obtained from Numerical Linked Cluster Expansion. I find excellent agreement between the two methods, as well as rapid convergence of the diagrammatic series. 
I also report results for the momentum distribution and kinetic energy of the electrons. 
\end{abstract}
\maketitle

\subsection{Introduction}
The Hubbard model \cite{Hubbard238} occupies a central position in condensed matter physics. Originally conceived as a description of Mott insulators, it is now primarily associated with cuprate superconductivity -- Specifically, it is believed to qualitatively describe electronic properties of the two-dimensional copper-oxide layers where superconductivity arises \cite{PWA}. But perhaps as important, the Hubbard model allows an elementary realisation of one of the central challenges in physics today; Accurate analysis and characterisation of macroscopic fermionic systems with strong interactions.
Correspondingly, it has been studied with a broad spectrum of techniques, ranging from Dynamic Mean Field Theory or its cluster generalisations \cite{PhysRevLett.62.324,PhysRevB.75.045118,PhysRevB.77.033101,PhysRevLett.106.047004,PhysRevLett.110.216405,RevModPhys.77.1027}
to 
wave function methods \cite{PhysRevLett.87.217002,PhysRevB.95.024506}, the Density Matrix Renormalisation Group theory \cite{PhysRevLett.69.2863}, Numerical Linked Cluster Expansion \cite{PhysRevLett.97.187202,PhysRevE.75.061119} and Auxiliary-field Quantum Monte Carlo \cite{PhysRevB.39.839,PhysRevLett.62.591,PhysRevB.40.506,afqmc}. It has also been addressed experimentally with optically trapped ultra cold atomic gases \cite{PhysRevLett.115.260401,PhysRevLett.116.175301,nature14223}.
Yet despite considerable progress, a reliable phase diagram in the macroscopic limit has still not been produced to this day, and frequently different numerical protocols produce results with notable discrepancies \cite{2006cond.mat.10710S,PhysRevX.5.041041}. 

The difficulties facing simulation of strongly correlated systems are partially rooted in the fact that the states emerging from these systems can posses spatial orders that make them unsuitable to mean field theory or cluster descriptions \cite{Nature375-561,GranularCuprates}.  
Another problem is the prospect of competing states that are very narrowly spaced in free energy, making uncontrolled approximations potentially very misleading \cite{Dagotto257,2006cond.mat.10710S}. 
A highly illustrative phenomena is the giant proximity effect, where superconductivity can be proximity induced in cuprate layers with a thickness of a hundred coherence lengths \cite{PhysRevLett.93.157002} -- Accurate theoretical representation of such a system is likely to be very delicate. 

In light of this it is clear that a reliable phase diagram for the Hubbard model is dependent on simulation techniques that can address the macroscopic system, and to do so with controllable and small error bars. 
A prominent method that fulfils these requirements is Bold Diagrammatic Monte Carlo (BDMC), which is based on stochastic sampling of skeleton graphs  of the diagrammatic expansion \cite{PhysRevLett.99.250201}. Combined with semi-analytic treatment of the weak instability in the Cooper channel, this technique was used to produce highly accurate and reliable data about the pairing symmetry in the case of moderate onsite repulsion ($U/t\le 4$) and filling factors of ($\langle n\rangle \le ~70\%$) \cite{0295-5075-110-5-57001}. 

It should be noted though, that $U/t\le 4$ does not correspond to the strongly correlated regime, especially not when doping the system this far away from half filling, and indeed, direct treatment of that parameter region with diagrammatic techniques is not tractable due to the presence of a large expansion parameter ($U$), as well as  increasingly large corrections to the density matrix from expansion terms when the doping is small \cite{0953-8984-29-38-385602}. 
A number of analytical techniques have however been forwarded in response to this. 

Universal fermionisation makes it possible to encode restrictions in the Hilbert space by the inclusion of non-hermitian projection terms in the Hamiltonian that remove certain states from the partition function. This further allows for second fermionisation, where doubly occupied sites are removed from the trace, and then reintroduced in the form of hard core bosons that are subsequently fermionised. The result of this is that the large expansion parameter $~U$ becomes a single particle energy in the new representation \cite{PhysRevB.84.073102}.

Spin-charge transformation can remove the increasingly large corrections to the density matrix at small doping, and does so by mapping the lattice fermions onto spins and spin-less fermions \cite{0953-8984-29-38-385602}. The spins can then be mapped onto fermions via Popov-Fedotov fermionisation \cite{JETP.67.535}. The result of this treatment is a fermionic theory where both expansion parameters and corrections to the density matrix remain moderate regardless of the parameters of the original model. 

While Popov-Fedotov fermionisation has been used together with BDMC to address frustrated spin systems \cite{PhysRevB.87.024407,PhysRevLett.110.070601,PhysRevLett.116.177203}, the remainder of these techniques remain untested. 
This article will discuss explicitly how to employ spin-charge transformation with bold diagrammatic Monte Carlo (SCT-BDMC) to address the 2D Hubbard model, and conduct a benchmark against results obtained via Numerical Linked Cluster Expansion (NLCE) for the case when $U=\infty$.

\subsection{Spin-charge transformation}
It is possible to map two-component lattice fermion Hamiltonians like the Hubbard and t-J models onto a description in terms of spins and spin-less fermions \cite{0953-8984-29-38-385602}. 
Under this transformation, the state vectors of the original description map according to
 \bea\nonumber
|\downarrow\uparrow\rangle  \to |n_\Delta=1\rangle\times |D\rangle,\;\;\;
|0\rangle  \to |n_\Delta=1\rangle\times |H\rangle,\;\\
\;|\uparrow\rangle\to |n_\Delta=0\rangle\times |\uparrow\rangle,\;\;\;
|\downarrow\rangle \to   |n_\Delta=0\rangle \times |\downarrow\rangle\;\label{statemap}.
\eea
Here, $n_\Delta$ is an occupation number for a spin-less fermion while the second part of the state describes a spin-$\frac{1}{2}$ degree of freedom. $|D\rangle$ and $|H\rangle$ can in principe be any superpositions of the form $\alpha|\uparrow\rangle+\beta|\downarrow\rangle $, provided that they are orthogonal, $\langle D|H\rangle=0$, but it is practical to take them to be eigenstates of $S_z$, for example by choosing $|D\rangle=|\uparrow\rangle,\;|H\rangle=|\downarrow\rangle$. 

A mapping of the original fermionic operators can then be constructed according to
\bea\nonumber
c_{i\sigma}^\dagger \to \Delta_i|\sigma,i\rangle\langle H,i|+\sigma\Delta_i^\dagger|D,i\rangle\langle \bar{\sigma},i|\\
c_{i\sigma} \to \Delta_i^\dagger|H,i\rangle\langle \sigma,i|+\sigma\Delta_i|\bar{\sigma},i\rangle\langle D,i|,\label{SCT}
\eea
where $\sigma=\pm 1$, $\bar{\sigma}=-\sigma$, $|\sigma=1\rangle=|\uparrow\rangle$ and $|\sigma=-1\rangle=|\downarrow\rangle$. 
Here, the terms in brackets operate on the spin degrees of freedom, while $\Delta,\Delta^\dagger$ annihilate or create a spin-less fermion. 

With the construction (\ref{statemap}), there is a one-to-one correspondence between the original degrees of freedom and the new representation, meaning that there is no requirement to enforce restrictions on the Hilbert space. Also, the operators defined in (\ref{SCT}) preserve anti-commutation relations, so that the mapping is exact. 

The original Hubbard model takes the form 
\bea
H=-\sum_{ \langle i,j\rangle,\sigma}t  c^\dagger_{i\sigma} c_{j\sigma}  
+\sum_i (U n_{\uparrow,i}n_{\downarrow,i} -\mu n_{i}), 
\label{classicHubb}
\eea
where $n_i=n_{\uparrow,i}+n_{\downarrow,i} =c^\dagger_{i\uparrow}c_{i\uparrow}+c^\dagger_{i\downarrow}c_{i\downarrow}$. 
The spin-charge transformed model is obtained by inserting  (\ref{SCT}) into (\ref{classicHubb}).
For the hopping term we obtain
\bea
 c_{i\sigma}^\dagger c_{j\sigma}\to
-\Delta_j^\dagger \Delta_i |\sigma,i\rangle\langle H,i|\times |H,j\rangle \langle \sigma,j| \label{holeprop}\\
+\Delta_i^\dagger\Delta_j |D,i\rangle\langle \bar{\sigma},i| \times |\bar{\sigma},j\rangle \langle D,j| \label{doubprop}\\
+\sigma \Delta_i\Delta_j |\sigma,i\rangle \langle H,i| \times |\bar{\sigma},j\rangle\langle D,j|\label{annihilation}\\
+\sigma  \Delta_i^\dagger \Delta_j^\dagger |D,i\rangle \langle \bar{\sigma},i |\times | H,j\rangle \langle \sigma,j|.\label{creation}
\eea
This term is thus decomposed into four parts, describing the propagation of a hole (\ref{holeprop}), propagation of a doublon (\ref{doubprop}) as well as annihilation (\ref{annihilation}) and creation (\ref{creation}) of a doublon-hole pair. 

The single site terms map according to
\bea\sum_i (U n_{\uparrow,i}n_{\downarrow,i} -\mu n_{i})\to\\
\sum_{i}\Big\{\mu \Delta_i^\dagger\Delta_i +(U-2\mu)\Delta_i^\dagger \Delta_i |D,i\rangle \langle D,i|\Big\} \label{singleSite}.
\eea
Thus, we obtain a chemical potential for the spin-less fermions which is $-\mu$, as well as an interaction term $\sim (U-2\mu)$ which gives the energy difference between a doublon and a hole. 

Since spins cannot be treated directly with diagrammatic techniques, it is necessary to map them onto fermions. This can be achieved via Popov-Fedotov fermionisation \cite{JETP.67.535}. The state vectors then transform according to
\bea
|S=\sigma\rangle \to |n_\sigma=1,n_{\bar{\sigma}}=0\rangle,\label{FPstateVectors}
\eea
where $n_{\sigma}$ is an occupation number for a spin-$\sigma$ fermion and $\bar{\sigma}=-\sigma$. Thus, a given spin is represented by a fermion carrying the very same spin. The spin-operators are then expressed in fermionic operators according to
\bea
|\sigma,i\rangle \langle \sigma,i| = \Big(\frac{1}{2}+\sigma S_{i,z}\Big),\;|\sigma,i\rangle \langle \bar{\sigma},i| =  a^\dagger_{i,\sigma} a_{i,\bar{\sigma}},\;\label{FPOperators}
\eea
where $S_z=(a^\dagger_{\uparrow}a_{\uparrow}-a^\dagger_{\downarrow}a_{\downarrow})/2$.
An important point here is that (\ref{FPstateVectors}) introduces unphysical states with zero or two spin-carrying fermions that have no corresponding spin state.
These can however be removed from the partition function by the inclusion of an imaginary chemical potential for the spin-carrying fermions of the form $\frac{ i\pi}{2\beta}(n_{i\sigma}+n_{i\bar{\sigma}}-1)$. This term does not affect physical states that have precisely one such fermion, but ascribes a phase to unphysical states with zero or two spin-carrying fermions so that they cancel from the partition \cite{JETP.67.535}. 

We can now write down a fully fermionic dual representation of the Hubbard model on the form
\bea
H\nonumber
=t\sum_{\langle ij\rangle ,\sigma}\Big\{\Delta_j^\dagger \Delta_i |\sigma,i\rangle\langle H,i|\times |H,j\rangle \langle \sigma,j| \\ \nonumber
-\Delta_i^\dagger\Delta_j |D,i\rangle\langle \bar{\sigma},i| \times |\bar{\sigma},j\rangle \langle D,j| 
\\ \nonumber
-\sigma \Delta_i\Delta_j |\sigma,i\rangle \langle H,i| \times |\bar{\sigma},j\rangle\langle D,j|
\\\nonumber
-\sigma  \Delta_i^\dagger \Delta_j^\dagger |D,i\rangle \langle \bar{\sigma},i |\times | H,j\rangle \langle \sigma,j|\Big\}
\\\nonumber
+\sum_{i}\Big\{\mu \Delta_i^\dagger\Delta_i +(U-2\mu)\Delta_i^\dagger \Delta_i |D,i\rangle \langle D,i|\\
+\frac{ i\pi}{2\beta}(n_{i\sigma}+n_{i\bar{\sigma}}-1)\Big\}\label{SCT_Hubbard}
\eea
where the spin operators are defined according to (\ref{FPOperators}).
In the new model, the large expansion parameter $\sim U$, which describes onsite repulsion has been replaced with a smaller term $\sim U-2\mu$, which gives the energy difference between a doublon and a hole. The rest of the interaction energy is relegated to the chemical potential of the spin-less fermions. In particular, this term vanishes at half-filling.

A natural comparison for the spin-charge transformation is the slave-boson technique \cite{0305-4608-6-7-018,PhysRevLett.57.1362}. However an important difference is that the introduction of a spin, rather than a Boson results in a direct correspondence between states in the respective representations (\ref{statemap}). This eliminates the need for restrictions on the Hilbert space that arise in the slave-boson approach, greatly simplifying diagrammatic treatment. 
The absence of these restrictions is a property that it shares with a number of closely related dual mappings that have been used in the context of mean-field calculations and other analytical approaches. 
In particular, applying the transformation (\ref{SCT}) to the t-J model (see \cite{0953-8984-29-38-385602}) exactly reproduces the dual representation considered by Khaliullin in \cite{1990JETPL..52..389K}. Likewise, a similar transformation for the Hubbard model was proposed by \"Ostlund and Granath \cite{PhysRevLett.96.066404}.
It should be stressed however that these works do not employ fermionisation and consequently do not derive fermionic models that can be addressed with diagrammatic techniques.

The best way to solve (\ref{SCT_Hubbard}) is dependent on model parameters. 
If $U-2\mu$ is sufficiently small, then this model can be treated directly with SCT-BDMC as there are no large parameters in the problem. 
When $U\gg t$, then it may be motivated to approximate the Hubbard model with the  t-J model \cite{PhysRevB.18.3453} which is defined on a restricted Hilbert space -- This treatment is exact in the case when $U=\infty$, and corresponds to $J=0$.
The most challenging situation occurs when $U-2\mu$ is large, but we cannot motivate using the t-J model as an approximation. In this scenario we have to employ second fermionisation \cite{PhysRevB.84.073102}, which involves projecting out the doublons via universal fermionisation, and then reintroducing them in the form of hard core bosons that are then fermionised. The result of this treatment is that the interaction term $\sim U-2\mu$ becomes a single particle energy of a fictitious fermion, albeit at the expense of a very complicated theory.  

This work will focus on the case when $U=\infty$, though the generalisation to the t-J model is straightforward, see \cite{0953-8984-29-38-385602}. 
In this scenario, doublons have infinite energy, and can therefore be regarded as forbidden, meaning that the Hubbard model effectively becomes a system of noninteracting particles on a restricted Hilbert space. Thus, we can remove three out of four kinetic terms in (\ref{SCT_Hubbard}) that describe propagation of a doublon as well as creation/annihilation of a doublon-hole pair, and only keep hole propagation. Likewise, we can remove the term $\sim U-2\mu$ which gives a correction to the doublon energy. We thus obtain a model of the form
\bea
H\nonumber
=t\sum_{\langle ij\rangle,\sigma}\Delta_j^\dagger \Delta_i |\sigma,i\rangle\langle H,i|\times |H,j\rangle \langle \sigma,j| 
\\
+\Big\{\sum_{i}\mu \Delta_i^\dagger\Delta_i
+\frac{ i\pi}{2\beta}(n_{i\sigma}+n_{i\bar{\sigma}}-1)\Big\}
\label{Restricted},
\eea
which is understood to be defined on a restricted Hilbert space where doublons are forbidden.  
It should be stressed that when $U=\infty$, this treatment involves no amount of approximation. 

To address this model with diagrammatic methods, the restriction on the Hilbert space must be explicitly encoded in the Hamiltonian, and this can be achieved through universal fermionisation \cite{PhysRevB.84.073102}. 
This technique requires the addition of two elements in the Hamiltonian: First, a field of {\it auxiliary} fermions, and secondly, an imaginary coupling between this field and the forbidden states in the Hilbert space:
\bea
H'=H+\sum_j\frac{i\pi}{\beta}P_j (n_{j,A}-1/2),\;\;[H,P_j]=0 \label{universal}.
\eea
Here, $n_{j,A}=c_{jA}^\dagger c_{jA}$ is a number operator for the auxiliary fermions. Meanwhile, $P_j$ is a projection term onto the forbidden subspace which returns $1$ if the site $j$ has a forbidden configuration, and $0$ otherwise. 
Since the prohibited states correspond to doubly occupied sites in this problem, it follows that the projection term takes the form of a number operator for doublons, i.e., $P_j=n_{j,\uparrow}n_{j,\downarrow}$, in the original representation (\ref{classicHubb}). 
Under spin-charge transformation this maps onto $P_j=\Delta_j^\dagger \Delta_j |D,j\rangle \langle D,j|$. It is also clear that $P_j$ commutes with $H$, which contains only projected hopping which will not create, destroy or move doublons. 

The effect of the construction (\ref{universal}) is that doublons are removed from the partition function. 
 Specifically, if acting on a state where the site $j$ is doubly occupied we find $P_j=1$, thus summing over $n_{j,A}=0,1$ we obtain contributions to the partition function with opposite sign that cancel.
 	However, if $P_j=0$, they enter with the same sign. Hence, computing the trace we obtain 
\bea
Z'=\Tr e^{-\beta H'}=2^N\Tr_{\Omega} e^{-\beta H}
\eea
where $\Omega$ is the physical/allowed subspace and $N$ is the number of lattice sites. This means that expectation values can be computed in the restricted Hilbert space according to
\bea
\langle \hat{O}\rangle =\frac{1}{Z'}\Tr e^{-\beta H'} \hat{O}.\label{expectation}
\eea
Choosing $|D\rangle=|\uparrow\rangle$, $|H\rangle=|\downarrow\rangle$ in the transformation (\ref{SCT}) and inserting the fermionised spin-operators (\ref{FPOperators}) into the Hamiltonian (\ref{Restricted},\ref{universal}) we can write down an explicit Hamiltonian on the form
 \bea
H
=\sum_{\langle ij\rangle}t\Big\{
 \Delta_j^\dagger \Delta_i\Big(\frac{1}{2}-S_{i,z}\Big) \Big(\frac{1}{2}-S_{j,z}\Big)\label{sdown}
 \\
+\Delta_j^\dagger \Delta_i\Big(a_{i,\uparrow}^\dagger a_{i,\downarrow}\Big)\Big(a_{j,\downarrow}^\dagger a_{j,\uparrow}\Big)\Big\}  \label{sup}
 \\ 
+\sum_{i}\Big\{\mu \Delta_i^\dagger\Delta_i +\frac{i\pi}{\beta}(n_{i,A}-\frac{1}{2}) \Delta_i^\dagger \Delta_i \Big(\frac{1}{2}+S_{i,z}\Big) \\
+\frac{ i\pi}{2\beta}(n_{i\sigma}+n_{i\bar{\sigma}}-1)\Big\},\;\;\;S_z=\frac{1}{2}(a_\uparrow^\dagger a_\uparrow-a_\downarrow^\dagger a_\downarrow)
\label{SCTHUBBARD}.
\eea
Formally this model is not Hermitian as it possesses two imaginary projections terms. However, these are only nonzero when acting on forbidden states, so (\ref{SCTHUBBARD}) is a symmetric operator in the restricted Hilbert space where there are no doubly occupied sites, and where every site has precisely one spin-carrying fermion defined by (\ref{FPstateVectors}). 

Although the model (\ref{SCTHUBBARD}) is more complicated than the original Hubbard model, it is free of two properties that are detrimental to diagrammatic treatment -- It does not feature a large expansion coefficient ($\sim U$), and it does not exhibit large corrections to the density matrix from nonlinear terms (see \cite{0953-8984-29-38-385602} for a discussion). Thus, despite the fact that we are considering a system with infinite onsite repulsion, it is tractable to diagrammatic treatment.

\begin{figure}[!htb]
\includegraphics[width=\linewidth]{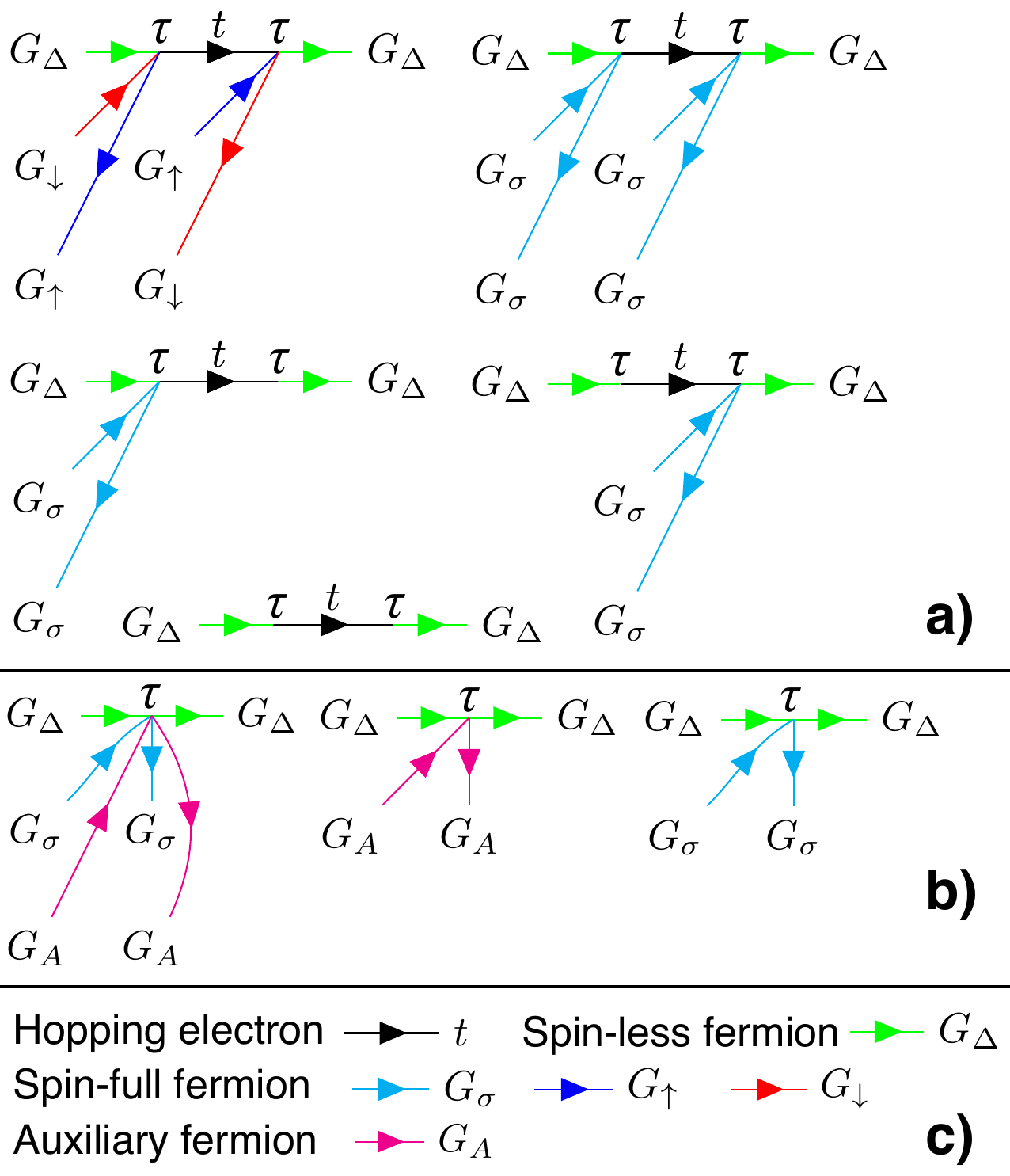}
\caption{
The elementary diagrammatic objects that emerge from the model (\ref{SCTHUBBARD}): (a) Terms originating from hopping electrons, (b) Projection terms used to enforce the restricted Hilbert space. The line types are described in (c).
This work uses a description based on imaginary time ($\tau$) and momentum ($\bar{k}$), though it possible to work with any combination of real-space/momentum and imaginary time / Matsubara frequencies. Thus, all lines are functions of $(\tau_2-\tau_1,\bar{k})$.
}
\label{vertices}
\end{figure}

Expansion of (\ref{SCTHUBBARD}) in a diagrammatic series gives rise to the set of elementary vertices shown in Fig. \ref{vertices}. Here, (a) corresponds to electron propagation while (b) gives the projection terms $\sim\frac{i\pi}{\beta}$.

\subsection{Diagrammatic treatment}
Sampling of the diagrammatic series of (\ref{SCTHUBBARD}) can be achieved with a worm algorithm as outlined in for example \cite{PhysRevB.87.024407,DiagramVsEmulator}.
The basis of this approach is a Metropolis type random walk in the space of diagram topologies and their internal variables. By the introduction of a fictitious diagrammatic element in the form of the {\it worm}, ergodicity can be achieved with a relatively simple set of updates. 
For the purpose of normalisation, one also adds an additional state to the space of diagrams, namely the {\it normalisation sector}. 

\begin{figure}[!ht]
\includegraphics[width=\linewidth]{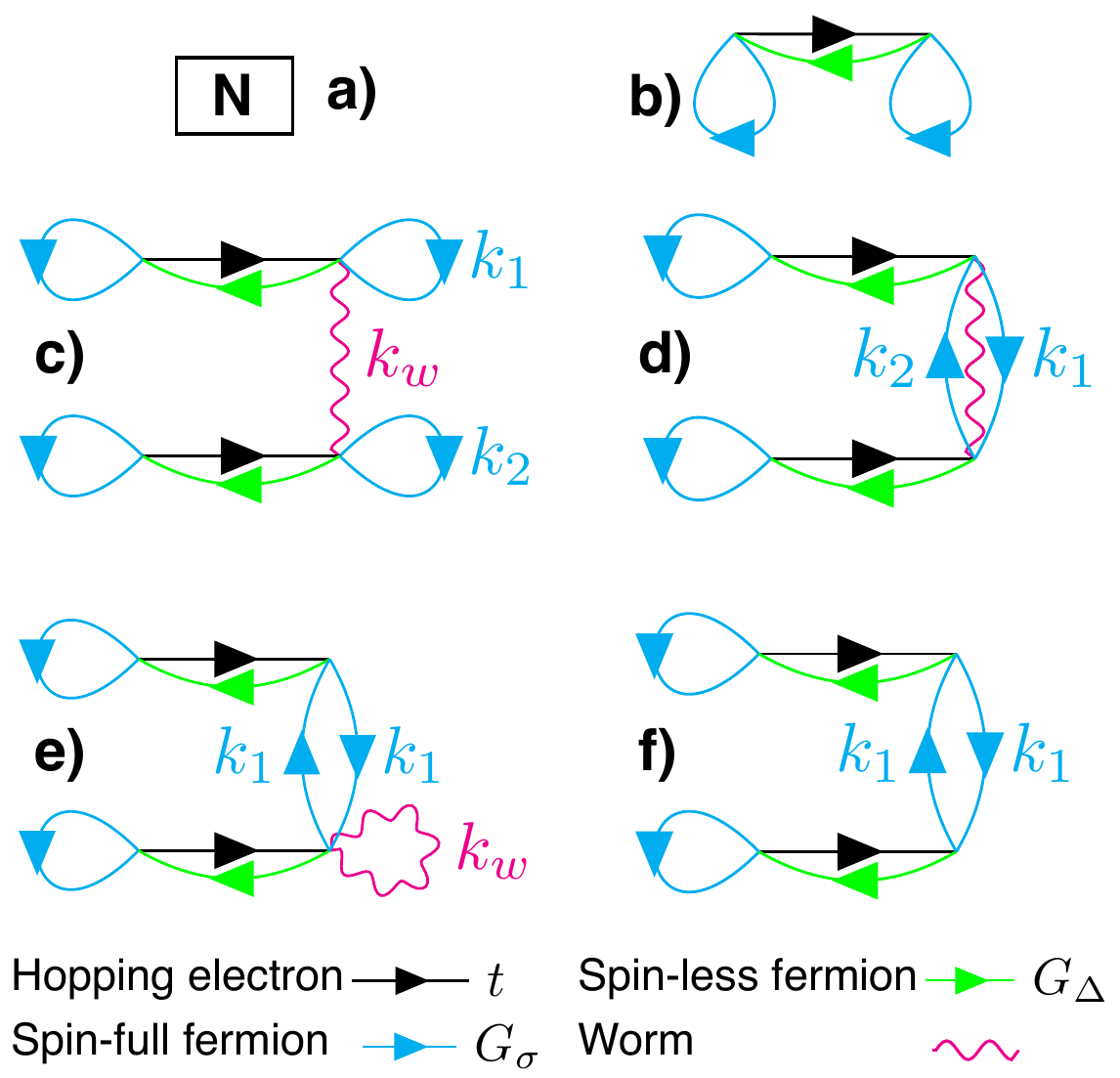}
\caption{
Basic updates: (N) is an artificial configuration that exists outside the diagrammatic space and is used for normalisation. It is connected to the set of first order diagrams through  leave/enter the normalisation sector (a $\Leftrightarrow$ b).
Add/remove vertex (b $\Leftrightarrow$ c) changes the diagram order by one and either creates or destroys a worm. Commute operators (c $\Leftrightarrow$ d) changes the topology while preserving the diagram order. The worm absorbs the momentum difference $k_w=k_2-k_1$ between the commuted particle lines so that momentum is preserved at every vertex. Moving one of the worm ends (d $\Leftrightarrow$ e) along a particle line changes its momentum by $k_w$. Add/remove worm (e $\Leftrightarrow$ f) connects configurations where a worm closes on itself with a worm-free state. Finally we allow translating a vertex or pair of vertices in imaginary time (not depicted).
}
\label{updates}
\end{figure}

The set of updates used in this work is shown in Fig. \ref{updates}. 
The normalisation sector is ascribed a weight $W$ (taken to be $W=1$ is this work). Through {\it leave/enter the normalisation sector} (a $\Leftrightarrow$ b) it is connected to the set of first order diagrams. 
Diagrams of a higher order are reached via {\it add/remove vertex} (b $\Leftrightarrow$ c). This update will take an element from Fig, \ref{vertices}, contract all the particle lines, and attach it to an already existing vertex via a worm with momentum $k_w=0$ (or remove such an object). {\it Commute operators} (c $\Leftrightarrow$ d) changes the operators being contracted, and thus the topology of the diagram. This can only be applied to particle lines that share vertices with the worm. The difference in momentum between the lines $k_2-k_1$ is absorbed by the worm, and thus momentum is preserved at every vertex. In {\it move worm} (d $\Leftrightarrow$ e), one of the worm ends moves to another vertex via an interaction/particle line, and in doing so, changes its momentum by $\pm k_w$ so that momentum remains preserved at all vertices. Once the worm closes on itself, it can be removed (e $\Leftrightarrow$ f) (or likewise, we can introduce a worm that closes on itself). Finally, time translation changes the imaginary time of a vertex or pair of vertices. 

The probability of accepting an update $A\leftrightarrow B$ is given by
\bea
\frac{P^{\text{Accept}}_{A\to B}}{P^{\text{Accept}}_{B\to A}} =\Big[\frac{P^{\text{Try}}_{A\to B}}{P^{\text{Try}}_{B\to A}}\Big]^{-1} \frac{W(A)}{W(B)}\label{Accept}
\eea
where $P^{\text{Try}}_{A\to B}$ is the probability of proposing the update $A\to B$, and $W(A)$ is the weight of the configuration $A$, which is given by
\bea
W(A)=|\Pi_n G_n(\tau_n,\bar{k}_n)\Pi_m t(\bar{k}_m)|.\label{Weight}
\eea
Thus, the probability of generating a topology and set of internal variables is proportional to the absolute value of that configuration. 

  As the worm is an unphysical object, it follows that the diagrams (\ref{updates}, c-e) are not part of the actual diagrammatic expansion of the model (\ref{SCTHUBBARD}) and correspondingly, they should not give any contribution to the observable being measured. Rather, they should be thought of as unphysical states that are introduced to simplify the construction of an ergodic Metropolis type process. 
 
\begin{figure}[!ht]
\includegraphics[width=\linewidth]{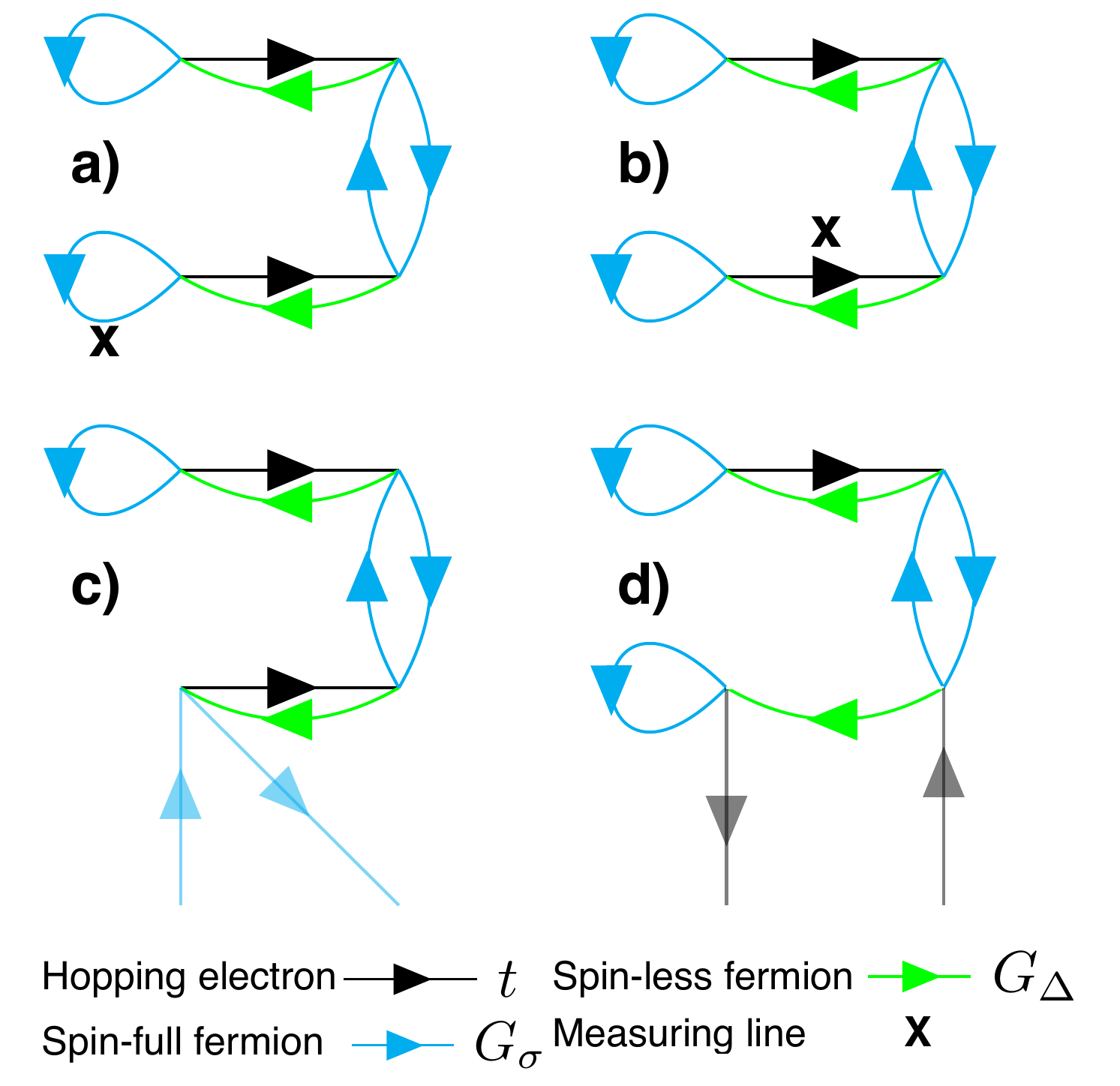}
\caption{
Schematic view of the measurement process. 
A measuring line is chosen stochastically. In (a) a spin-full fermion    (marked by an X) is selected and so the measuring line is given by $G_\sigma(\tau,\bar{k})$. The tagged line is removed (b), and the diagram is interpreted as a contribution to the self energy $\Sigma_\sigma(-\tau,\bar{k})$. 
In (c), a hopping line is selected, and the interpretation (d) is a contribution to the polarisation operator $\Pi_t (-\tau,\bar{k})$.
If a measurement is attempted while in the normalisation sector, then a number is added to a special normalisation bin (denoted "N" in Fig. \ref{updates}). The properly normalised self-energy or polarisation operator is obtained by dividing by this bin.  
}
\label{mlines}
\end{figure}

Since the diagrams in Fig. \ref{updates} are elements of the partition function, they do not immediately give any observables. This is instead achieved via the use of a stochastically chosen measuring line as shown in Fig. \ref{mlines}.
If a particle propagator $G_\sigma$ is chosen as in (\ref{mlines},a) then this is interpreted as a contribution to the self energy $\Sigma_\sigma(\tau,\bar{k})$ (\ref{mlines},b). Likewise, if a hopping line is selected (\ref{mlines},c), then the interpretation is a contribution to the polarisation operator $\Pi_t(\tau,\bar{k})$ (\ref{mlines},d). 
Only one measuring line is employed at a given time.
If a measurement is made while in the normalisation sector (Fig. \ref{updates}, a), then a data point is added to the normalisation bin. 

The reconstruction of the self-energy/polarisation operator from the sampled data is then achieved as follows:
First we note that all diagram configurations are connected to the normalisation sector via Metropolis type updates, with acceptance probabilities given by \ref{Accept}. This implies that the probability-density of generating the configuration $A$ is related to the probability of reaching the normalisation sector ($N$) according to
\bea
\frac{P(A)}{P(N)}=\frac{W(A)}{W(N)}
\eea
where we recall that $W(N)=1$. 
Thus, the magnitude of a contribution is implicitly taken into account by the Metropolis process. However, since $P(N)$ is generally unknown a priori, the results must be normalised at the end of the simulation. 
When conducting a measurement with a measuring line say $G_\sigma(\tau_m, \bar{k}_m)$, we obtain a contribution to $\Sigma_\sigma$ at the point $(-\tau_m, \bar{k}_m)$ with a phase $\phi$ determined by the configuration sampled. In more conventional theories the diagrams generally only exhibit a sign so that $\phi=0$ or $\pi$, but due to the fermionisation techniques employed here, the Hamiltonian possesses imaginary terms, meaning that the integrand can take complex values.
Now, the self-energy must be reconstructed from a set of discrete events distributed in momentum and imaginary time. It is for this purpose convenient to use an orthonormal function space: 
\bea
\Sigma_\sigma(\tau,\bar{k})=\sum_{ijk} c_{ijk}\chi_i(\tau)\xi_{jk}(\bar{k}).
\eea
The (un-normalised) contributions to the coefficients $c_{ijk}$ then correspond to projections of these events onto the function space:
\bea
\delta c_{ijk} = \langle e^{i\phi}\delta(\tau-\tau_m)\delta(\bar{k}-\bar{k}_m) | \chi_i(\tau)\xi_{jk}(\bar{k}) \rangle. 
\eea
At the end of the simulation the results are normalised according to
\bea
c_{ijk}\to \frac{c_{ijk}}{n(N)}
\eea
where $n(N)$ is the number of times a measurement was conducted in the normalisation sector. 
For an extensive discussion of detail balance and normalisation in Diagrammatic Monte Carlo, see \cite{PhysRevB.87.024407}.
 
\subsection{Observables}
The full Greens function and dressed hopping-line can be obtained from the self energy and polarisation operator respectively via Dyson type equations of the form:
\bea
G_\sigma (\omega,\bar{k})=\frac{1}{1/G_\sigma^0(\omega,\bar{k})- \Sigma_\sigma(\omega,\bar{k})}\label{dressedG}\\
\kappa_t (\omega,\bar{k})=\frac{1}{1/t(\bar{k})- \Pi_t(\omega,\bar{k})},\label{dressedInt}
\eea
where $\Sigma,\;\Pi$ denote the {\it proper} self energy and polarisation operators that only contain irreducible diagrams, while $\omega$ are (discrete, imaginary) Matsubara frequencies. 

In Diagrammatic Monte Carlo, the filling factor and momentum distribution is generally obtained from the Greens function, $n_\sigma(k)=G_\sigma(\tau\to -0,\bar{k})$. However because of the transformation above, the basic degrees of freedom are fictitious particles and it is therefore more practical to obtain the properties of the electrons in the original model from the polarisation operator corresponding to the hopping-line. 
Specifically, we can write down a projected Greens function for holes in the restricted Hilbert space according to
\bea\nonumber
\tilde{G}_{h,\sigma}(x-x')=\\
\langle T_\tau [1-n_{e,\bar{\sigma}}(x')]c_{e,\sigma}(x')c_{e,\sigma}^\dagger(x)[1-n_{e,\bar{\sigma}}(x)]  \rangle, \label{GeDef}
\eea
where $x,x'$ are space-time coordinates and $\bar{\sigma}=-\sigma$. Applying the transformation (\ref{SCT}) to (\ref{GeDef}) we obtain 
\bea\nonumber
\tilde{G}_{h,\sigma}(x-x')\to \\
\langle T_\tau  \Delta^\dagger(x')\Delta(x) |\sigma,x\rangle\langle H,x|\times |H,x'\rangle \langle \sigma,x'|\rangle.
\eea 
This in turn is related to the full polarisation of the $t$--line according to
\bea\nonumber
\tilde{G}_{h,\sigma}(\omega,\bar{k})= \Pi^{full}_{t,\sigma}(\omega,\bar{k})\\\nonumber
=\Pi_{t,\sigma}(\omega,\bar{k})+\Pi_{t,\sigma}(\omega,\bar{k})t(\bar{k})\Pi_{t,\sigma}(\omega,\bar{k})+...\\
=\frac{1}{1/\Pi_{t,\sigma}(\omega,\bar{k})-t(\bar{k})},\label{Ge}
\eea
and so the projected Greens function can be directly related to the polarisation operator of the $t$--line. 

It should be stressed here that the object (\ref{GeDef}) differs from conventional Greens functions in that it contains projection operators associated with the restricted Hilbert space. Nonetheless it does give direct access to the momentum resolved hole density (and thus also the particle density) through 
\bea
n_h(\bar{k})=\tilde{G}_{h,\sigma}(\tau= -0,\bar{k}),& \;\;\;& \text{(Hole density)}\\
 n(\bar{k})=1-n_h(\bar{k})\label{nk}, & \;\;\;& \text{(Particle density)}\\
\rho=\int \frac{d\bar{k}}{(2\pi)^2} n(\bar{k}). &\;\;\; &\text{(Filling factor)}\label{filling}
\eea
In contrast, we generally have $\tilde{G}_{h,\sigma}(\tau= +0,\bar{k})\not=n_h(\bar{k})-1$, contrary to the case of ordinary fermionic Greens functions. 

It is in principe possible to compute arbitrary operator expectation values by constructing the corresponding diagrammatic elements and performing Monte Carlo sampling. For example, we could compute spin correlations by introducing an operator which in the original degrees of freedom takes the form
\bea
\hat{O}_{ij}=J_{ij}S_i \cdot S_j.
\eea
Using (\ref{SCT}) we obtain
\bea
\hat{O}_{ij}= J_{ij} (1-\Delta_i^\dagger \Delta_i)(1-\Delta_j^\dagger \Delta_j)S_i \cdot S_j.
\eea
Taking $J_{ij}$ as the measuring line (see Fig. \ref{mlines}) we can compute $\langle \hat{O}_{ij} \rangle$ for an interacting system. Since $\hat{O}$ is not part of the diagrammatic expansion, we need only include one such object to obtain the expectation value (but multiple orders in $t$). Correlators with more than two points can be computed in the same way, but require more complicated measuring objects than lines, and also give rise to more intricate data sets. 

\begin{figure}[!htb]
\includegraphics[width=\linewidth]{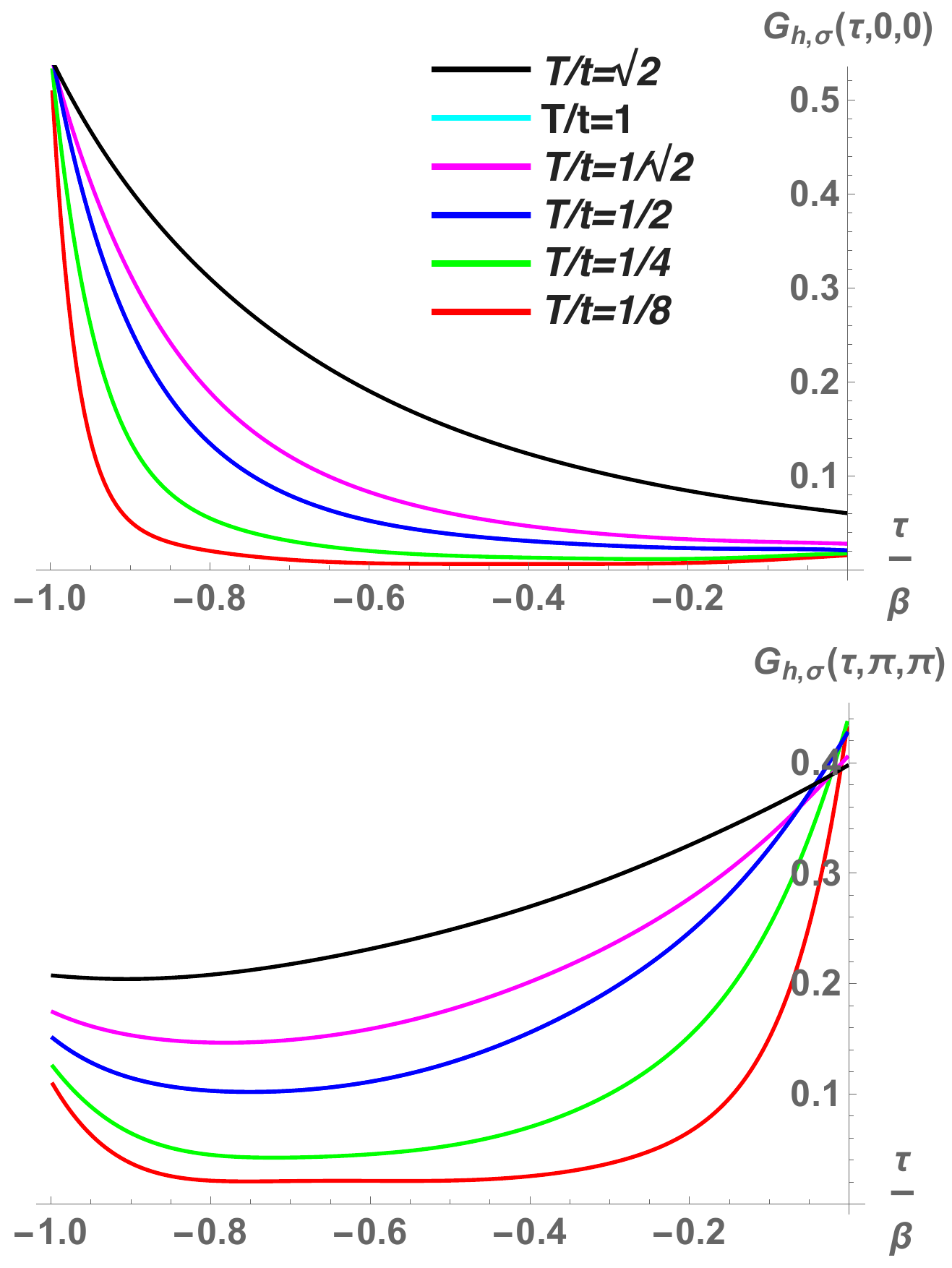}
\caption{
Projected Greens function for the holes, $\tilde{G}_{h,\sigma}(\tau,\bar{k})$ (Eg. \ref{Ge}) at the high-symmetry points $\bar{k}=\{0,0\}$ and $\bar{k}=\{\pi,\pi\}$. The model parameters are $\mu=2$, $U=\infty$ while the temperatures are shown in the insert. The expansion order is $O=3$ in all cases.   
}
\label{Green}
\end{figure}

\begin{figure*}[!htb]
 \hbox to \linewidth{ \hss
\includegraphics[width=\linewidth]{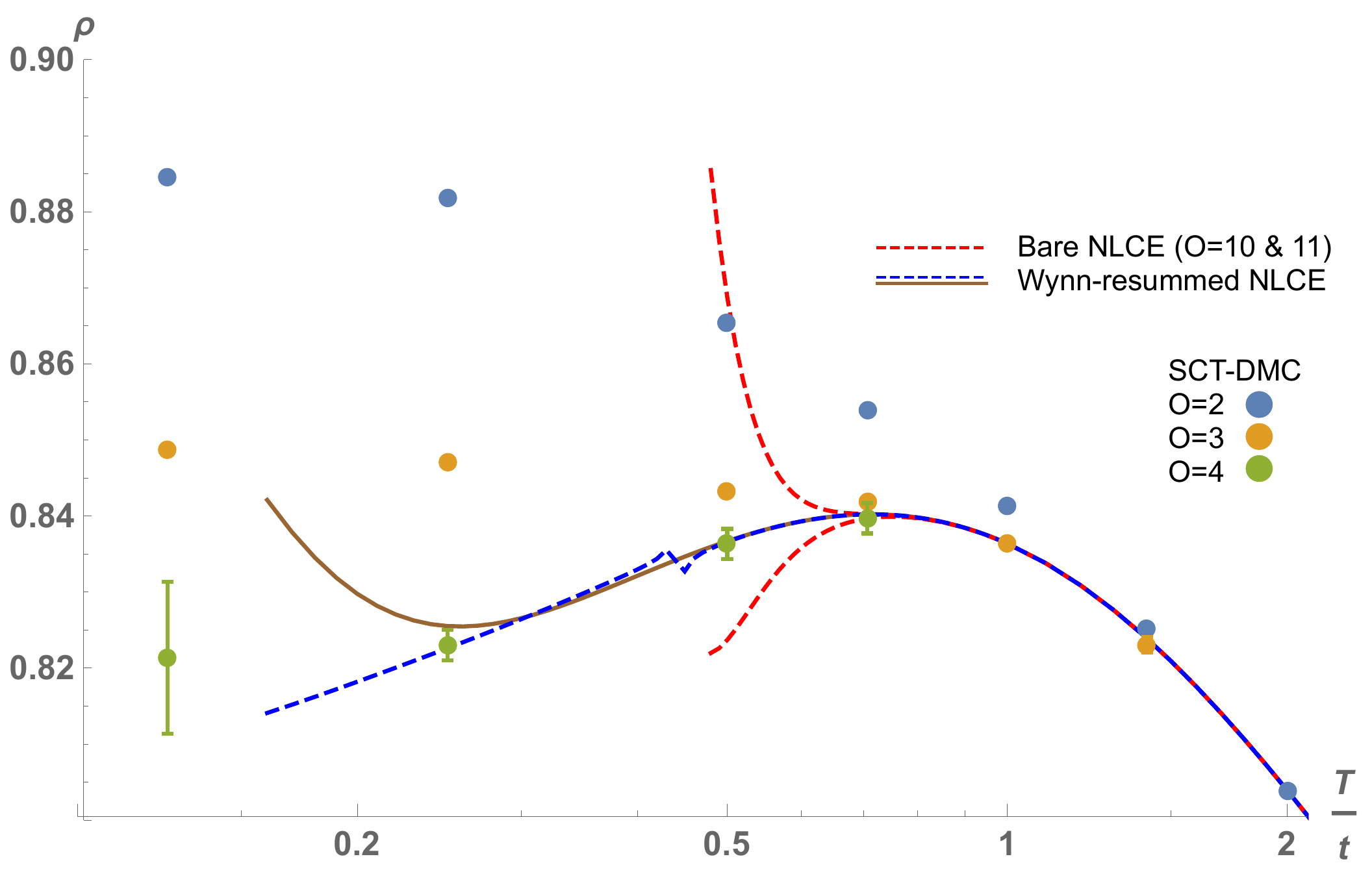}
 \hss}
\caption{
Filling factor (Eq. \ref{filling}), as a function of temperature (in units of hopping) for the 2D Hubbard model when $U=\infty$ and $\mu/t=2$. 
The coloured disks show results obtained through SCT-BDMC, with expansion orders ranging from $O=2$ to $O=4$. 
The solid and dashed lines show results obtained by Khatami et al. from Numerical Linked Cluster Expansion (NLCE) \cite{PhysRevE.89.063301}. Data is given both for bare expansions to orders $10$ \& $11$, as well as after Wynn resummation. 
At $T/t=2$ there is little discrepancy between techniques or expansion orders, indicating rapid convergence with either method. With decreasing temperature, bare NLCE and the lowest order expansions start to diverge.
Bare NLCE remains reliable down to $T/t\approx\sqrt{1/2}$, and these results are reproduced by SCT-BDMC within error bars. When $T/t<\sqrt{1/2}$, comparison can only be made to data obtained via Wynn resummation. 
In this region diagrammatic expansion to orders $3$ and $4$ start to diverge, indicating that higher order terms are required to see convergence. Still, expansion to fourth order gives results in good agreement with the resummed data. 
}
\label{results}
\end{figure*}

\subsection{Skeleton techniques}
It is possible to improve on naive sampling of the diagrammatic series by employing a {\it bold scheme} that exploits the repetitious nature of the diagrammatic expansion \cite{PhysRevLett.99.250201,PhysRevB.87.024407,2013arXiv1305.3901V}.
This work uses the perhaps most basic such approach, namely expansion in the dressed hopping line $\kappa_t(\tau,\bar{k})$ defined in (\ref{dressedInt}), rather than its bare counterpart $t(\bar{k})$. 
The result of this is that a certain class of topologies are accounted for to high, or even infinite order implicitly (since $\kappa_t(\tau,\bar{k})$ contains terms up to infinite order in $t$). So as to not register the same contributions twice, one must therefore omit any diagrams that belong to this class. 
Such {\it reducible} diagrams are easily identified, as they contain insertions from the polarisation operator. By definition, these can be cut loose from the remainder of the diagram by removing two $\kappa/t$--lines (since all topologies in $\Pi_t$ have one incoming and outgoing such line). By momentum conservation, these two lines have the same momentum, and correspondingly, it is sufficient to maintain a hash table of all line momenta and simply reject contributions that posses $\kappa/t$--lines with degenerate momenta. 
In the Hubbard model, skeleton techniques may be particularly important when $U/t\to \infty$, since this limit is associated with ferromagnetic correlations at low temperature, which translates to a large mean free path \cite{PhysRev.147.392,Motambaux,Andreev}.

\subsection{Simulation and benchmark}
To evaluate the Monte Carlo procedure outlined above, it is natural to compare it to results obtained with other state of the art numerical protocols. 
For the Hubbard model with infinite onsite repulsion, reliable data about the equation of state can be obtained via Numerical Linked Cluster Expansion (NLCE) \cite{TANG2013557,PhysRevE.89.063301}. 
This technique is exact in the limit of infinite cluster size, and correspondingly it is also controllable, as convergence of the result with respect to cluster size can be readily checked.

Currently published data is available for chemical potentials in the range $-2\le\mu/t\le2$ \cite{PhysRevE.89.063301}.
The method presented in this manuscript is aimed primarily at the case of small doping where the density of spin-less fermions is small. For this reason, comparison will be made for the case $\mu/t=2$, which results in filling factors in the range $0.8\le \langle n\rangle \le 0.85$ for the temperatures of interest. 

There is in principe nothing that prevents addressing systems with larger doping, but it should be expected that the smaller chemical potential results in a larger statistical weight of high order diagrams, since spin-less fermions mediate all processes, and this suggests that the sign error becomes more severe the further we are from half filling. Nonetheless, much of the interesting physics in the Hubbard model and indeed the cuprates occurs relatively close to half-filling -- Optimal doping for the high-temperature superconductors is generally less than $20\%$ for instance \cite{RevModPhys.78.17}.

The equation of state was obtained as follows: First, the polarisation operator $\Pi_t(\tau,\bar{k})$ was computed by sampling diagrams up to fourth order in the dressed hopping line $\kappa(\tau,\bar{k})$, for temperatures in the range $1/8\le T/t \le 2$.
The projected Greens function was then obtained using Eq. \ref{Ge}, see Fig. \ref{Green}. 
Finally, the filling factor was extracted using Eq. \ref{nk}, and plotted against data obtained via NLCE, see Fig. \ref{results}. 

Bare NLCE provides reliable results down to $T/t\approx 1/\sqrt{2}$, and these are also reproduced by SCT-BDMC simulations. 
In most of that temperature range, expansion to third order is sufficient to obtain accurate results for the filling factor, while there is a small correction at fourth order when $T/t= 1/\sqrt{2}$. 
Beyond this point, the diagrammatic expansion is not controllable since convergence could only be determined with access to higher order terms than $O=4$. 
Nonetheless, comparison of the fourth order expansion and Wynn-resummed data does show agreement within error bars, suggesting that the series is close to convergence even at lower temperatures while also attesting the resummation scheme in \cite{PhysRevE.89.063301}.

\begin{figure}[!htb]
\includegraphics[width=\linewidth]{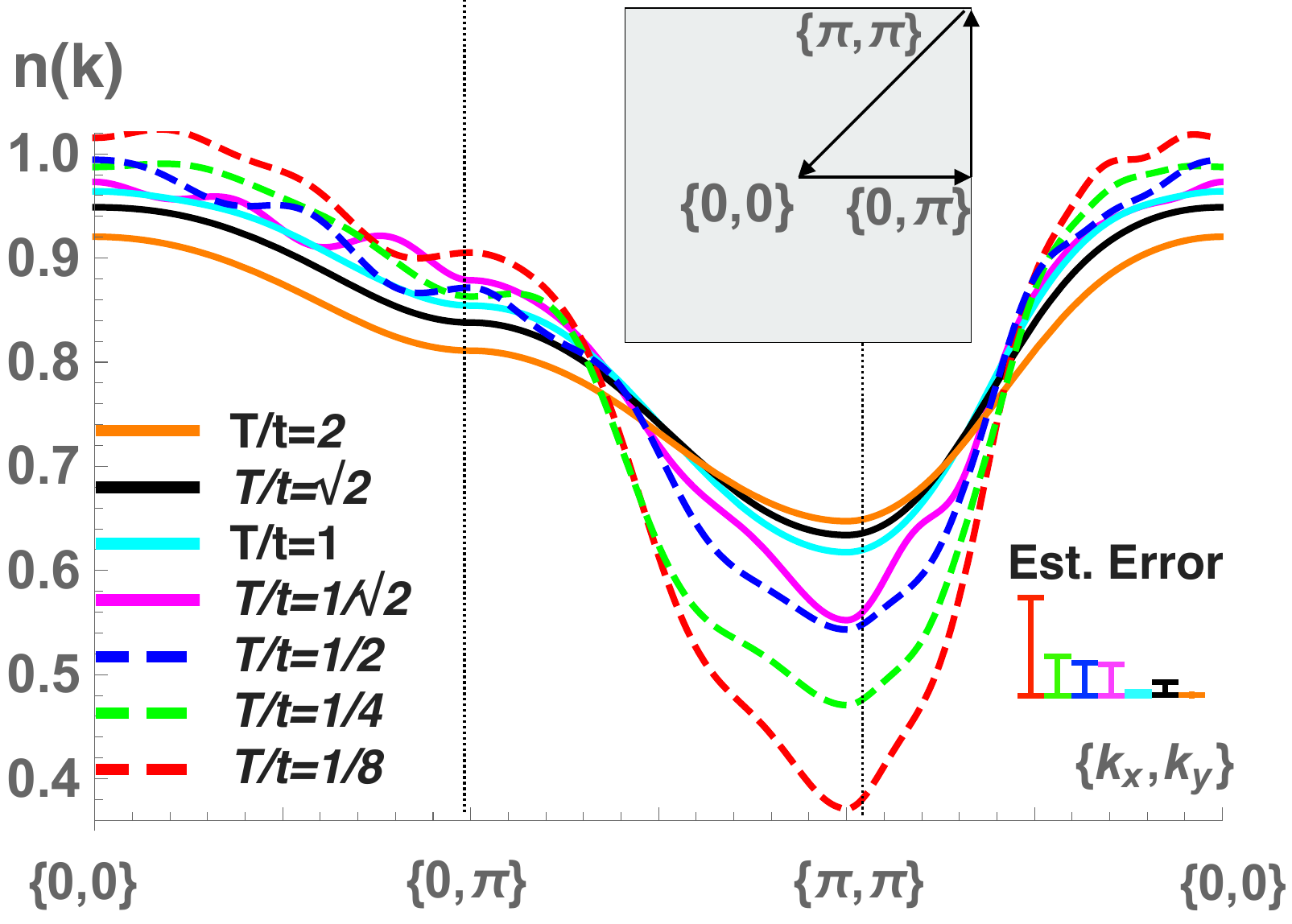}
\caption{
Electron density as a function of momentum ($n(k)$) obtained from Eq. \ref{nk}, at temperatures $1/8\le T/t \le 2$ and with $U=\infty,\;\mu=2t$, plotted along the high symmetry axes shown in the insert. The data corresponds to the highest expansion order shown in Fig. \ref{results} at every temperature. The error estimates were obtained by conducting multiple simulations and taking $\Delta n=\text{max}(|n_i(\bar{k})-n_j(\bar{k})|)$ for any $\bar{k}$, where $n_i,\;n_j$ are obtained from independent simulations. 
}
\label{rhoK}
\end{figure}

\begin{figure}[!htb]
\includegraphics[width=\linewidth]{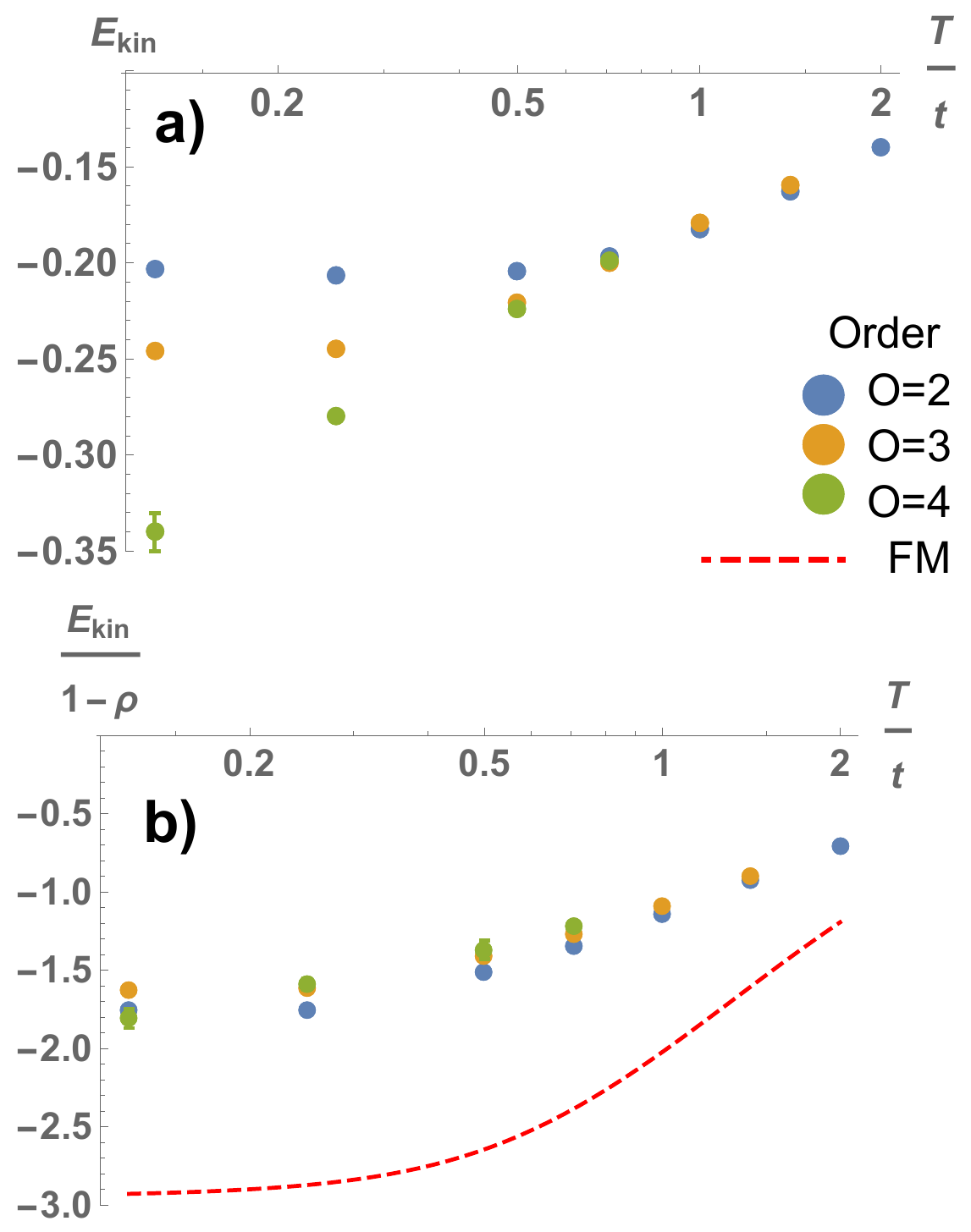}
\caption{
(a) Kinetic energy as function of temperature at expansion orders $2\ge O\ge 4$ for the case $\mu=2t, \; U=\infty$ (same as in Fig. \ref{results}). The two highest orders give results in good agreement down to $T=1/2$, i.e., slightly lower than the filling factor. At lower temperatures the correction from higher order diagrams is substantial.  
(b) Normalising by the carrier density we obtain the kinetic energy per hole which can be compared to the corresponding figure for an ideal ferromagnet, shown by the dashed line (see also Eq. \ref{FM}). The carrier density (number of holes per lattice site) is given by $1-\rho$ (see Eq. \ref{filling}).
}
\label{Ekin}
\end{figure}

The momentum dependence of the particle density is shown in Fig. \ref{rhoK}. At the lower temperatures there is an increasingly large drop in particle density centred around $\bar{k}=\{\pi,\pi\}$ which reaches below $40\%$ at $T=1/8$ (though further corrections beyond fourth order are likely). 

It is known from the Nagaoka theorem that the ground state of the Hubbard model with infinite repulsion and a single hole is ferromagnetic, with the hole being delocalised in the entire system \cite{PhysRev.147.392}. 
In the fully polarised environment, the hole behaves as an ideal lattice fermion. 
Once spin-fluctuations are present however, the density of states drops drastically close to the band edges \cite{brink}, and the motion is no longer ballistic \cite{PhysRevLett.116.247202,PhysRevB.96.014303}. 
Thus, the Nagaoka theorem can be understood from the context that an infinite mean-free path leads to the largest possible effective density of states at the band edges. 
Although the theorem only applies to a single hole, it is conceivable that ferromagnetism persists to finite, perhaps even substantial doping by this mechanism.
A natural comparison is therefore that of ballistic holes moving in an environment with infinite mean-free path:
\bea
H_{FM}=2t (\cos[k_x]+\cos[k_y])+\mu. \label{FM}
\eea
At $\bar{k}=\{\pi,\pi\}$ and $\mu=2$, this gives a hole density of $(1+e^{-2/T})^{-1}$,
 i.e., very close to unity at lower temperatures. Thus, the observed particle density in Fig. \ref{rhoK} suggests that the kinetic energy is significantly reduced due to spin fluctuations. 
This picture is also consistent with computations of the kinetic energy shown in Fig. \ref{Ekin}. Normalising the kinetic energy by carrier density and comparing to an ideal ferromagnet we find a difference of approximately unity in units of $t$ at the lower temperatures as a result of spin fluctuations (though the case $T=1/8$ has clearly not converged).

\subsection{Conclusions}
The starting point of this work is a conjecture about diagrammatic treatment of strongly correlated fermions that identifies two primary obstacles that prevent convergence of the series, namely a large expansion parameter as well as vast corrections to the density matrix.
Universal fermionisation and spin-charge transformation are two analytical techniques that have been forwarded to address these two respective issues and the aim of this work is correspondingly to test them in practice.

The main conclusion of this work is that via SCT-BDMC, based on a worm algorithm with boldification of the hopping line, it becomes possible to reproduce the equation of state for the Hubbard model at $U=\infty$. 
Crucially, this approach can also be generalised to other scenarios when the interactions are not infinite. 
Also, since this approach is based on standard diagrammatic methods, it can be combined with a wide range quantum field theory techniques. These include mean-field shifts / shifted action expansion \cite{PhysRevB.93.161102} to increase the radius of convergence, as well as more sophisticated boldification schemes. 
It also allows designing "custom" operators for the purpose of measuring, which gives access to essentially arbitrary observables. 

At the current stage the biggest deficit of the approach is a comparatively steep computational cost of higher order diagrams which limits the expansion to fourth order -- Beyond this point the sign error prevents extracting the polarisation operator with acceptable accuracy. This is a lower expansion order than what is typically achievable in diagrammatic simulations, and this can most likely be attributed to the complexity of the transformed model, especially the prevalence of three-body operators. 
Still, it becomes possible to resolve key features of the equation of state, in particular the drop in electron density below $T\approx \sqrt{1/2}$, even at this expansion order.
Thus, the results strongly support the initial conjecture and demonstrate that diagrammatic simulation techniques can be applied to strongly correlated systems and models with restricted Hilbert spaces given the right analytical techniques.

The current limitations in terms of expansion order can be overcome with more efficient sampling protocols.
Recently it was demonstrated that the series of connected diagram topologies can be expressed as a recursion of determinants, making it possible to use determinant methods in the macroscopic limit \cite{PhysRevLett.119.045701,0295-5075-118-1-10004}.
The result of this treatment is exponential rather than factorial scaling of the computational complexity with expansion order. Moreover, the dual model employed here exhibits particles that are non-dispersing, meaning that contractions of operators separated in space are known to be zero a priori, and this in turn allows the determinant to be decomposed into contributions from respective lattice points, dramatically reducing the computational effort (specifically, the complexity will grow by an exponent that is a fraction of that of a system of dispersing particles). 
Thus, the spin-charge transformed model does not only exhibit rapid convergence, but is also particularly well suited to the new sampling protocols based on determinants that are currently being developed. 

\subsection{Acknowledgements}
This work was supported by the Wenner-Gren Foundations in Stockholm and the Simons Collaboration on the Many Electron Problem. Computations were performed on resources provided by the Swedish National Infrastructure for Computing (SNIC) at the National Supercomputer Centre in Link\"oping, Sweden.
The author would like to thank Marcos Rigol for providing data for bench marking, as well as Boris Svistunov, Nikolay Prokof'ev and Riccardo Rossi for important input and discussions. 

\bibliography{biblio}

\end{document}